\documentclass[preprint]{aastex}
\usepackage[T1]{fontenc}
\usepackage{amsmath}
\usepackage[english]{babel}
\usepackage{graphics}

\begin{document}

\bibliographystyle{apj}
\slugcomment{Accepted by Geophysical and Astrophysical Fluid Dynamics}

\title{Shear and Mixing in Oscillatory Doubly Diffusive Convection}

\author{Francesco Paparella}
\affil{Department of Mathematics,
University of Lecce, Lecce, Italy}
\author{Edward A. Spiegel}
\affil{Astronomy Department, Columbia University,
New York, New York, USA}
\author{Suzanne Talon}
\affil{CERCA and D\'epartement de physique, Universit\'e de
Montr\'eal, Montr\'eal, Canada}



\begin{abstract}
To investigate the mechanism of mixing in oscillatory
doubly diffusive (ODD) convection, we truncate the
horizontal modal expansion of the Boussinesq equations
to obtain a simplified model of the process.  In the
astrophysically interesting case with low Prandtl number,
large-scale shears are generated as in ordinary thermal
convection.  The  interplay between the shear and the oscillatory
convection produces intermittent overturning of the fluid
with significant mixing.  By contrast, in the parameter
regime appropriate to sea water, large-scale flows are not
generated by the convection.  However, if such flows are imposed
externally, intermittent overturning with enhanced mixing is observed.
\end{abstract}


\newpage

\section{Introduction}

Doubly diffusive convection occurs when two material properties, such as
heat and salt with differing diffusion rates, affect the fluid density
and are stratified so that one is stabilizing and the other destabilizing
(Turner, 1979). When the stably stratified property diffuses more
rapidly, the onset of the convection is by way of a monotonic growth that gives
rise to structures known as \emph{salt fingers}. This kind of convection
is an effective mixer, the more so because it induces large scale shears
(Paparella \& Spiegel, 1999, hereafter PS) 
in the same way that ordinary Rayleigh-B\'{e}nard
convection does (Krishnamurti \& Howard, 1981, Howard \& Krishnamurti, 1986). 
On the other hand,
when the destabilizing constituent is diffused more rapidly, the onset of
convection is through growing oscillations.  The mechanism of mixing in this
case, that we shall refer to as oscillatory doubly diffusive, or ODD,
convection is less well understood than in the salt finger case and this
is the subject of the present work.

The question of mixing by ODD convection is not academic since this
form of doubly diffusive convection does arise in natural circumstances.
One example occurs in the cores of stars in late stages of their lives
(Spiegel, 1969). In the version of this process called semiconvection by
astronomers, the stabilizing constituent is a relatively heavy element,
such as helium in hydrogen-rich stellar material.  In many cases of
interest, the resulting stabilizing gradient of molecular weight opposes
an unstable entropy gradient. The problem that this raises is to determine
the rate of mixing of the heavy element outward in the star.  Though this
process involves some complications peculiar to the astrophysical
situation that we shall address in another place, we may usefully confine
ourselves here to the exploration of the basic fluid dynamical aspects
of mixing by ODD convection.  Suffice it to say that if the mixing caused
by semiconvection is effective, the convection can be self-regulating and
this is an important ingredient in estimating the speed at which the star
evolves.

Other natural occurrences of ODD convection are found in the Earth's
oceans, most notably, below the polar ice caps. There melting ice
releases cold fresh water above warmer saltier water and produces the
kind of situation we shall study here (Neal et al., 1969, Jacobs et
al., 1981). Another natural occurrence of ODD convection is seen in
\emph{meddies}, vortices of warm and salty Mediterranean water common
at mid depths in the eastern Atlantic (Ruddick, 1992). Intrusive
staircases present an alternance of fingering and
ODD stratification (Marmorino, 1991). Here too, there are
additional complicating features that we shall not confront in
studying the conventional plane-parallel situation stratified
(initially) only in the vertical direction. This ideal situation is in
fact the object of extensive experiments that have been carried out in
binary mixtures (Kolodner et al., 1990).  These have revealed much
about the nature of weakly nonlinear behavior of ODD convection.
Though the waves seen in these experiments will surely do some mixing,
we want to argue here that, at higher Rayleigh numbers, we may expect
to see induced shears like those seen by Krishnamurti \& Howard
(1981) in thermal convection.  Those shears, in concert with the
oscillatory convective motions, should lead to rapid mixing of
constituents.

To bring out these results, we shall use the same modal procedure that
we used in the monotonic case previously (PS).
This permits surveys of the parameter space with good
qualitative and semiquantitative results without the excessive demands
on numerical resources that are necessary for full simulations with
large aspect ratio. Modal truncations are known to yield results which
depend on the number of modes retained. This is illustrated, for
example, by Franceschini \& Tebaldi (1985) for relatively high
dimensional truncations of the Navier-Stokes equations. However, when
the focus is more on physical mechanisms than on the details of the
bifurcation diagram, even crude truncations have proven to be a
useful guide in the exploration of the dynamics of the underlying PDE.
Already in the simpler case of thermal convection, it was shown by
Howard and Krishnamurti (1986) that a model consisting of a small
number of ODEs captures some features of the large-scale shearing
flow.  In the particular case of ODD convection, a Galerkin approach
was used by Veronis (1965), to show oscillatory solutions in a system
of 5 ODEs (see also Da Costa et al., 1981). In a later study, Gough
\& Toomre (1982) used a truncation only in the horizontal direction,
as had been used in the purely thermal problem (Gough et al., 1975),
but they did not allow for a large-scale horizontal flow.  Their
results focus on the stability of an initial step-like density
stratification and they omitted the horizontal shear of the
Krishnamurti-Howard studies.

In our previous paper on the salt-finger case, we used modal truncation
in the horizontal direction to argue for the importance of the
horizontal flows for transport of heat and salt. Here we carry out an
analogous study for the ODD case.  In doing this, we would stress
that a Galerkin study such as ours is an exploratory tool and is intended
as a guide in rapidly identifying those features of the process that will
then deserve a more realistic and resource-consuming study by means of
direct numerical simulations.

Our plan is to present the equations in the next section in full and
truncated form. We restrict our attention here to two-dimensional
convection and refer to our previous paper (PS) for the
 derivation
of the truncated equations, which at this stage are the same for both
cases. In section 3 we explore the effect of a self-excited large-scale
flow on ODD convection with parameters that allow for an analogy with
the stellar case. Section 4 contains results for sea water parameters.
We discuss and summarize our findings in section 5.

\section{The equations}

We consider the Boussinesq equations for double diffusion in two
dimensions.  The velocity field can be computed using a stream
function \( \psi  \), for the horizontal and vertical velocities
\( u =\partial _{z}\psi \) and \(w = - \partial _{x}\psi \).
The basic equations are:
\begin{equation}
\label{2Dboussinesq}
\begin{array}{c}
\partial _{t}\nabla ^{2}\psi -J(\psi ,
\nabla ^{2}\psi )=-\sigma R_{T}\partial _{x}T+\sigma R_{S}\partial
_{x}S+\sigma \nabla ^{4}\psi \\
\partial _{t}T-J(\psi ,T)=\nabla ^{2}T\\
\partial _{t}S-J(\psi ,S)=\frac{1}{\tau }\nabla ^{2}S
\end{array}
\end{equation}
 where the Rayleigh numbers are
\begin{equation}
R_{T}=\frac{g\alpha \Delta Td^{3}}{\kappa _{T}\nu }
\qquad {\rm and} \qquad R_{S}=\frac{g\beta \Delta Sd^{3}}{\kappa
_{T}\nu } \; ; \label{Ray}
\end{equation}
\( \alpha \) and \( \beta \) are respectively the thermal and saline
expansion coefficients, \( g \) is the acceleration of gravity,
\( \nu \) is the viscosity and \( \kappa _{T} \) is the thermal
diffusivity. We also use the Prandtl number \( \sigma =
\nu /\kappa_{T}\) and the Lewis number \( \tau =\kappa _{T}/\kappa _{S}
\), where \( \kappa _{S} \) is the diffusivity of salt (or other solute).
The equations have been made dimensionless with a unit of length, \( d \),
where the height of the fluid layer is \( \pi d \), and a unit of time
\( d^{2}/\kappa _{T} \); the imposed temperature and salinity
differences across the layer \( \Delta T \) and \( \Delta S \) serve as
units for temperature and salt concentration. The signs of the Rayleigh
numbers determine the particular kind of doubly-diffusive instability that
triggers the motion. Here our interest is limited to the oscillatory case,
which requires that the salinity difference across the fluid layer be
stabilizing and that the temperature difference be destabilizing.  In our
notation, this is achieved when both Rayleigh numbers are
negative.

In this study we consider only the case with an initially
stable density stratification, that is \( R_{T}>R_{S} \), where
instability sets in through growing oscillations.  (We recall also
that oscillatory instability is possible for \( R_{S}>R_{T} \),
Baines \& Gill, 1969.) As in PS, we use a modal version
 of these
equations. We split the variables into a horizontally averaged part and a
fluctuating part: \( \psi =\overline{\psi }(z,t)+\psi '(x,z,t) \);
\(T=\overline{T}(z,t)+T'(x,z,t) \); \( S=\overline{S}(z,t)+S'(x,z,t) \).
The fluctuating part is expanded as a Fourier sum along the \( x \)
direction, with amplitudes depending on \( (z,t) \). The expansions are
then plugged back into the equations (\ref{2Dboussinesq}), and
truncated by retaining only the horizontally averaged quantities, and
the amplitudes associated with a single horizontal wavenumber
\( k_h \).

Our model (modal) equations are these nine p.d.e.s:

\begin{equation}
\label{eq:1mode_model}
\begin{array}{c}
\partial _{t}U=\partial _{z}\left( \sigma
\partial _{z}U+\frac{1}{k_h }W^{+}\partial _{z}W^{-}-\frac{1}{k_h
}W^{-}\partial _{z}W^{+}\right) \\
\partial _{t}\overline{T}=\partial _{z}
\left( \partial _{z}\overline{T}-W^{+}\Theta ^{+}-W^{-}\Theta ^{-}\right)
\\
\partial _{t}\overline{S}=\partial _{z}
\left( \frac{1}{\tau }\partial _{z}\overline{S}-W^{+}\Phi ^{+}-W^{-}\Phi
^{-}\right) \\
\partial _{t}\Theta ^{+}=\mathcal{D}
\Theta ^{+}+k_h U\Theta ^{-}-W^{+}\partial _{z}\overline{T}\\
\partial _{t}\Theta ^{-}=\mathcal{D} \Theta ^{-}-k_h U\Theta
^{+}-W^{-}\partial _{z}\overline{T}\\ \partial _{t}\Phi ^{+}=\frac{1}{\tau
}\mathcal{D}\Phi ^{+}+k_h U\Phi ^{-}-W^{+}\partial _{z}\overline{S}\\
\partial _{t}\Phi ^{-}=\frac{1}{\tau }
\mathcal{D}\Phi ^{-}-k_h U\Phi ^{+}-W^{-}\partial _{z}\overline{S}\\
\partial _{t}\mathcal{D}W^{+}=\sigma
\mathcal{D}^{2}W^{+}-\sigma k_h ^{2}\left( R_{T}\Theta ^{+}-R_{S}\Phi
^{+}\right) -k_h W^{-}\partial _{zz}U+k_h U\mathcal{D}W^{-}\\
\partial _{t}\mathcal{D}W^{-}=\sigma
\mathcal{D}^{2}W^{-}-\sigma k_h ^{2}\left( R_{T}\Theta ^{-}-R_{S}\Phi
^{-}\right) +k_h W^{+}\partial _{zz}U-k_h U\mathcal{D}W^{+}
\end{array}
\end{equation}
Here \( \Theta ^{+} \), \( \Theta ^{-} \)are the amplitudes associated
with, respectively, the sine and the cosine modes of the temperature
expansion. Analogously \( \Phi ^{+} \), \( \Phi ^{-} \)describe
the salinity fluctuations. We use amplitudes \( W^{+} \), \( W^{-} \) for
the vertical velocity fluctuations, as well as the horizontal
velocity \( U(z,t) \), in place of the amplitudes of the modal expansion
for \( \psi ' \) and of \( \bar{\psi } \). Finally,
\( \mathcal{D} \equiv \partial _{zz}-k_h ^{2} \).
The large scale flow represented by \( U \) has often been neglected
in previous studies of oscillatory convection. However, it is now
experimentally well established that such a large scale, shearing flow
does spontaneously appear in Rayleigh-B\'enard convection (Siggia, 1994)
where it changes the heat transport scaling law; we will show
in the following the important role it may play in ODD convection.

\section{ODD convection at low Prandtl and high Lewis number.}

Conditions favorable to the onset of ODD convective instability
are thought to occur in the core of a large variety of stars. In the
simplest scenario, they are
related to the existence of a central convection zone, in which
hydrogen is slowly converted to helium. The mass fraction \( Y \) of
helium influences the differential buoyancy in the same way that the
concentration of salt does in thermohaline convection. In modern
evolutionary models,
this central convection zone retreats, leaving behind a stabilizing radial
helium gradient. In the case of massive stars, the entropy gradient in the same
region is typically unstable and this leads to a form of ODD convection
known in astrophysics as {\it semiconvection} because it is effective
in transporting helium but not heat.

In stars, the thermal diffusivity is orders of magnitude greater than
diffusivity of the helium.  Typical values for the Prandtl and the Lewis
numbers (Merryfield, 1995) are
\begin{equation} \sigma \sim 5\cdot 10^{-5},\, \, \tau \sim 10^{8}.
\end{equation}
While solving the equations for the stellar plasma in this parameter
regime is a challenge for the numerical modeler, insight into the
nature of the process may be gained by solving the modal equations
(\ref{eq:1mode_model}) at low Prandtl number and high Lewis number.
We would hope in this way to ascertain whether the overstable
conditions can lead to a state of overturning convection. This is
important in estimating the rate of mixing of the chemical species in
the doubly-diffusive region.  To get an insight on the nature of
mixing in such conditions, we shall adopt here the values \(
\sigma=10^{-2} \), and \( \tau=10^{2} \) and present results for \(
R_{T}=-200 \), \( R_{S}=-400 \). Those calculations have been performed
using a resolution of 64 modes in $z$. We choose the horizontal wavenumber
\( k_h =1 \), which is close to the wavenumber of maximum growth rate
in the linear theory for that set of parameters. (For a discussion of
the linear stability analysis of doubly diffusive convection
appropriate for the boundary conditions that we use here, see Baines
and Gill, 1969.)  We have explored other parameter regimes than those
described here and found a rich and complicated range of behaviors.
Extremely long transients may precede the time-asymptotic states.  We
have tried to choose representative situations in the regime with low
Prandtl number and high Lewis number.  These display two basic
dynamical states: time-dependence with intense mixing (quantified
in section~\ref{sec:lagrangian}), and steady motion with no mixing.
Both are present within the parameter set that we next discuss in
detail.

\subsection{The wave-shear mixing cycle\label{sec:wave-shear}}

Equations (\ref{eq:1mode_model}) are solved with fixed temperature,
fixed salinity and free slip boundary conditions, using the numerical
code described in PS.
Initially we set \( \overline{T}=\overline{S}=-z \);
\( U=A\cos (z) \); all the other variables are set equal
to \( A\sin (z) \).

\begin{figure}
{\centering \resizebox*{1. \textwidth}{!}
{\rotatebox{-0}{\includegraphics{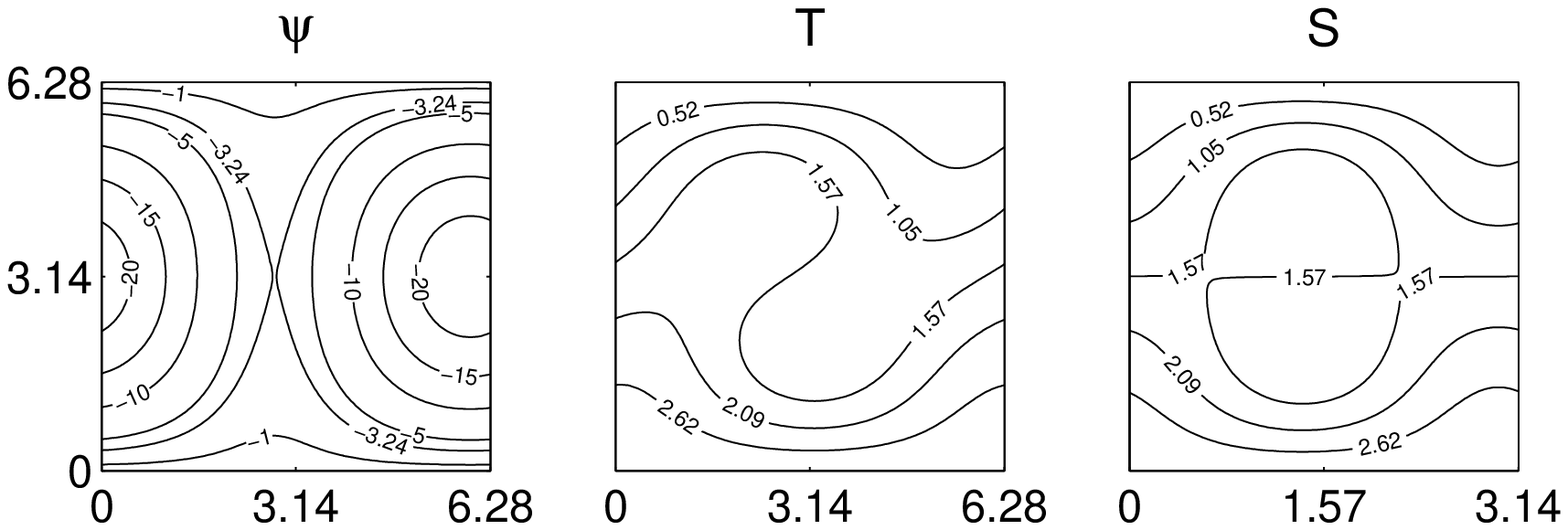}}} \par}

\caption{Stream function, temperature and salinity fields
of the steady solution for \protect\(\sigma=10^{-2}\protect \), 
\protect\(\tau=10^2\protect \), \protect\( R_T=-200 \protect \),
\protect\(R_S = -400\protect \). The fields are periodic in the horizontal. One
period is shown.
\label{fig:steady_smallA}}
\end{figure}

\begin{figure}
{\centering \resizebox*{0.7 \textwidth}{!}
{\rotatebox{-0}{\includegraphics{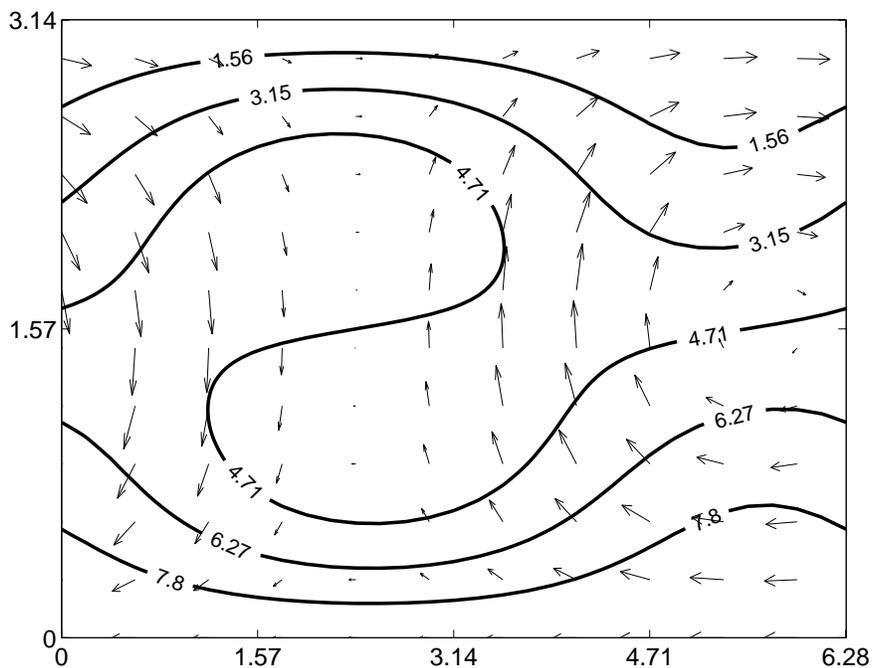}}} \par}

\caption{Density (solid line) and velocity field (arrows) for the flow
of fig.~\ref{fig:steady_smallA}. The density field is defined as:
$\rho=T+(R_S/R_T)S$. \label{fig:density_inversion}}
\end{figure}

When \( A \) is small (we used \( A=10^{-4} \)) the nonlinearities are
initially negligible, and the initial perturbation evolves quickly
into a good approximation to the fastest growing eigenfunction. As the
nonlinear terms become important, the system settles to a steady
solution, shown in figure~\ref{fig:steady_smallA}. This solution is
dominated by a large scale shearing velocity, sustained by a single
convective cell, which, in our example, circulates in the clockwise
direction. The reflected configuration with an anti-clockwise cell is
also possible and is obtained by starting with a negative
amplitude. The most remarkable feature of this solution is its steady,
stable density inversion, associated with a saddle point in the stream
function ({\em cf.} fig.~\ref{fig:density_inversion}). The formation of this
structure, an incomplete wave roll-up, is preceded by a temporal
development that begins in the linear stages of the growth with a pair
of cells of opposite signs, as in ordinary convection.  This motion
induces a wave pattern in the temperature and salinity fields. When
the amplitudes $W^+$ and $W^-$ reach the order of unity, the onset of
the large scale flow \( U \) makes the convective cells more and more
asymmetrical, advecting the temperature and salinity wave crests close
to each other as they encroach on the weaker cell, and stretching them
apart in the stronger one.  Finally, the weaker cell disappears
altogether, leaving a tongue of warm, salty fluid folded over a tongue
of cold, fresh fluid, above a saddle point in the stream function, in
a remarkable advective-diffusive equilibrium.  Further simulations
show that this is a stable solution.

We also performed a run where the absolute values of the Rayleigh numbers
were slowly and continuously lowered at the rate of $(1.5\cdot10^{-6})$
units per thermal time, while keeping their ratio constant. The steady
solution survives up to $R_T\approx -14.5$ when it abruptly falls to the
conductive state, supporting the supposition that it appears through a
sub-critical bifurcation.

\begin{figure}
{\centering \resizebox*{0.7 \textwidth}{!}
{\rotatebox{-0}{\includegraphics{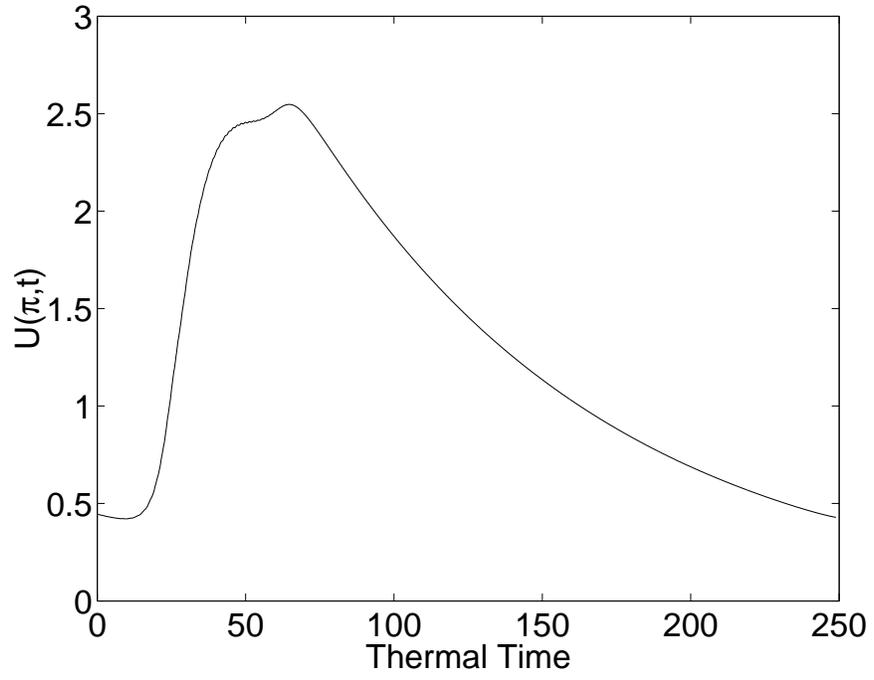}}} \par}

\caption{Time evolution of the shearing velocity $U$
at $z=\pi$.\label{fig:Uz0}}
\end{figure}

\begin{figure}
{\centering \resizebox*{0.72 \textwidth}{!}
{\rotatebox{-0}{\includegraphics{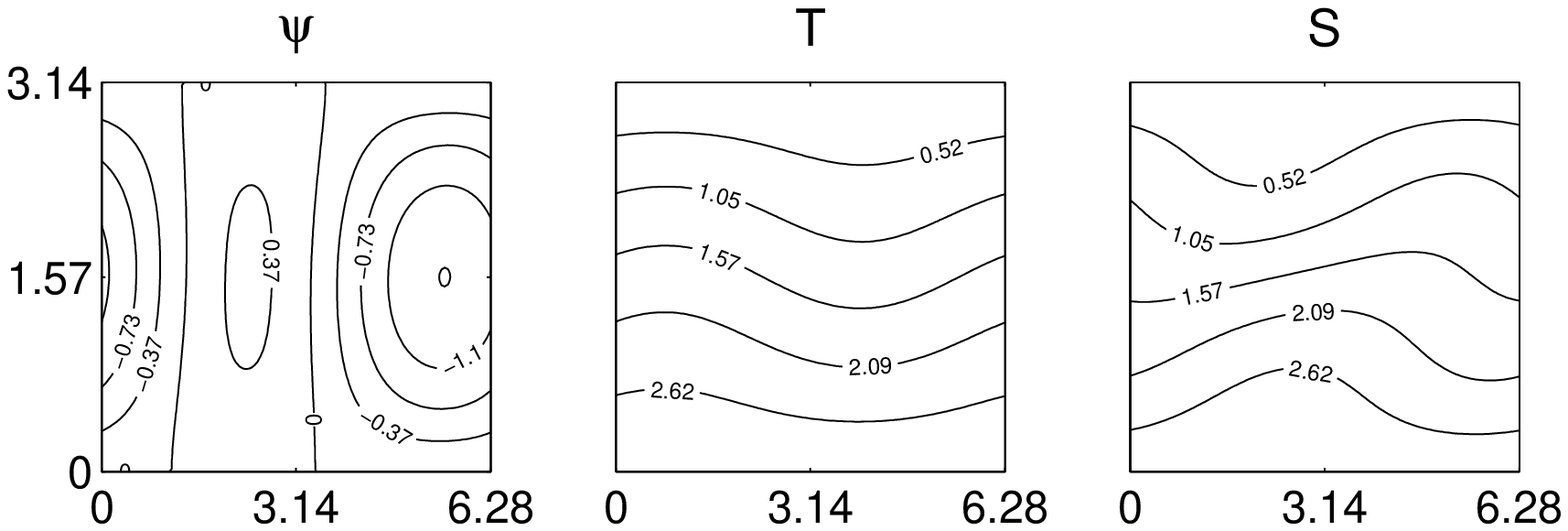}}} \par}

{\centering \resizebox*{0.72 \textwidth}{!}
{\rotatebox{-0}{\includegraphics{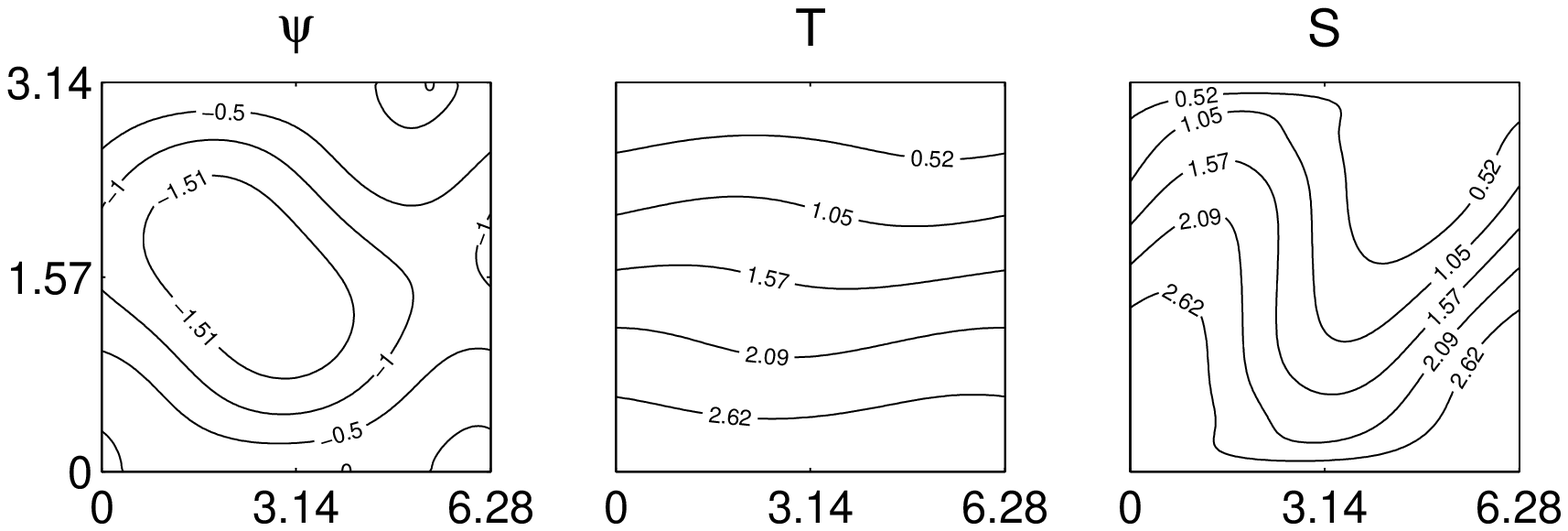}}} \par}

{\centering \resizebox*{0.72 \textwidth}{!}
{\rotatebox{-0}{\includegraphics{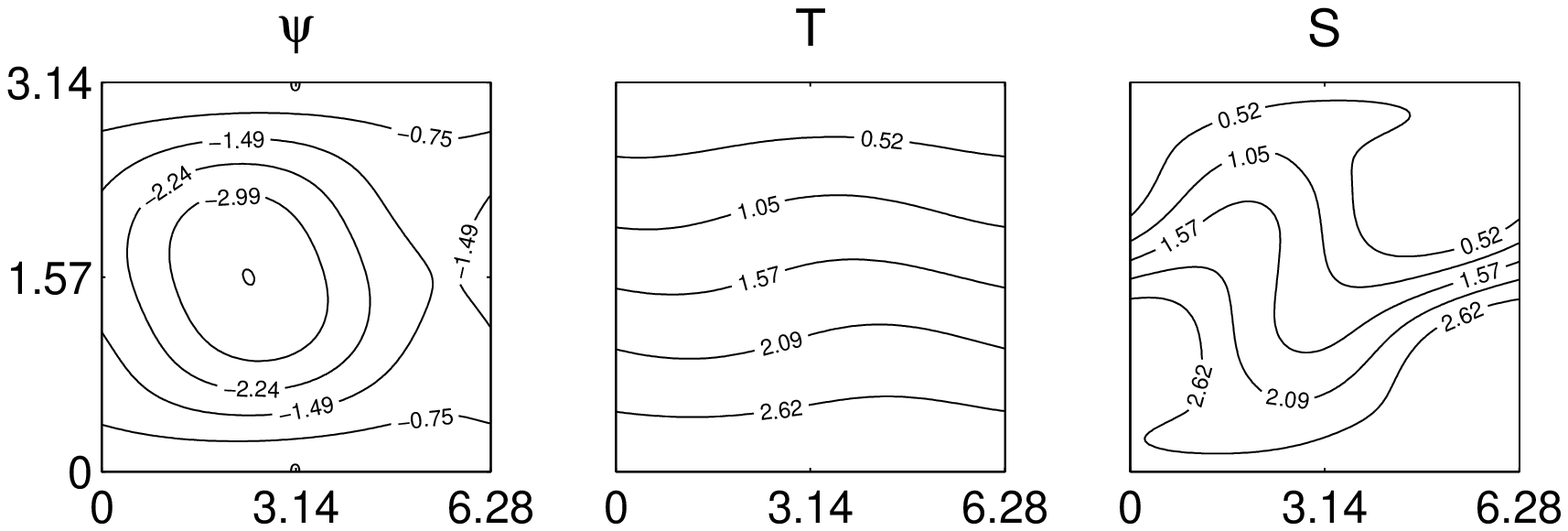}}} \par}

{\centering \resizebox*{0.72 \textwidth}{!}
{\rotatebox{-0}{\includegraphics{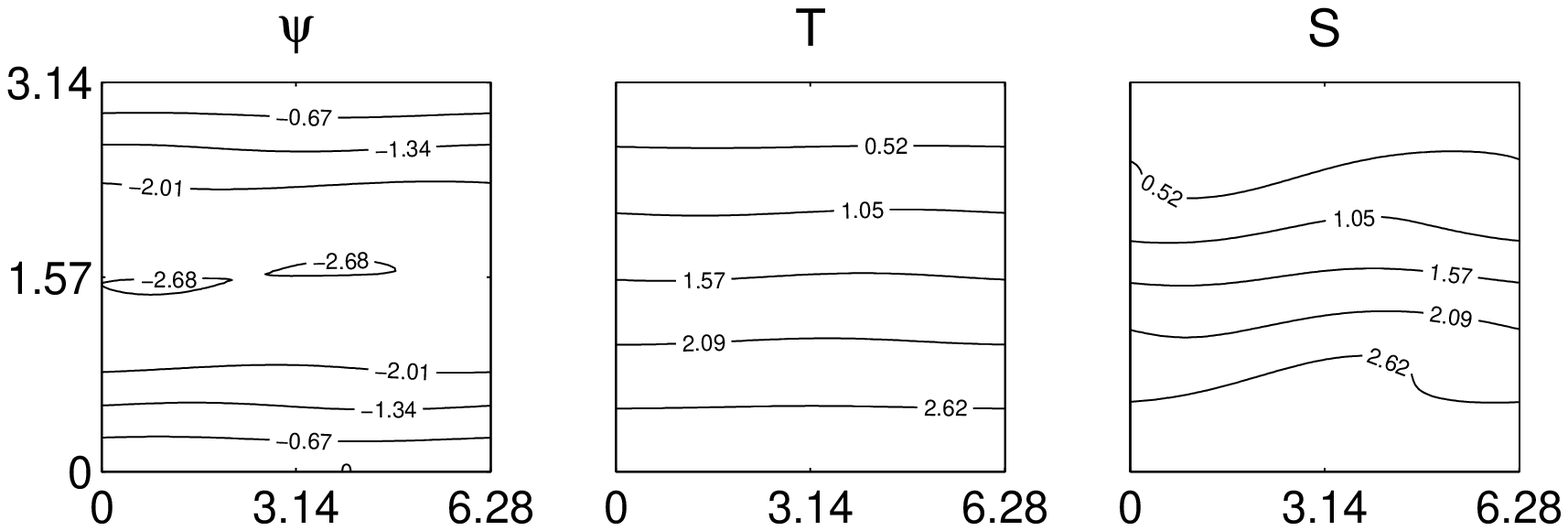}}} \par}

{\centering \resizebox*{0.72 \textwidth}{!}
{\rotatebox{-0}{\includegraphics{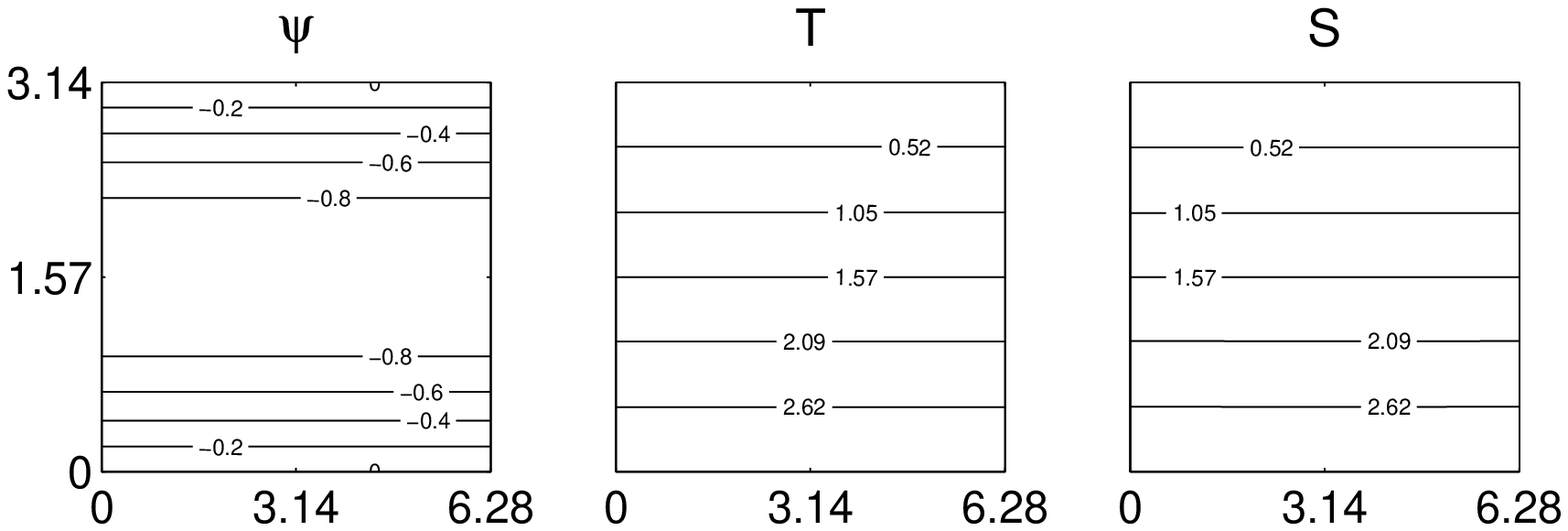}}} \par}

\caption{Stream function, temperature and salinity fields
during the cycle of Fig. \ref{fig:Uz0}. They correspond to the
times (from top to bottom): $t=14$, $t=24$, $t=48$, $t=63$, $t=163$.
The fields are periodic in the horizontal.
One period is shown. \label{fig:cycle}}
\end{figure}

A completely different dynamics appears if one sets $A=1$ in the
initial conditions. We then find a solution that proceeds in cycles of
approximately \( t_c \simeq 246 \) thermal times. In figure~\ref{fig:Uz0} we
show the time evolution of \( U \) at \( z=\pi \), during one such
cycle.  For convenience we set \( t=0 \) just before the large-scale
flow reaches its minimum amplitude. At this time, the stream function
shows a large, anti-clockwise cell, placed beside a smaller region of
weak clockwise circulation. Growing, wave-like patterns shape the
temperature and the salinity fields as they are advected across the
domain by the large-scale flow (fig.~\ref{fig:cycle}, first panel from
top). In this case, however, the shear is a leading order effect, and
slows the growth of the overstable oscillations.

Shortly thereafter, the asymmetry of the convective cells drives up
the amplitude of the large-scale flow. The second panel from the top
of figure~\ref{fig:cycle} shows the fields at time \( t=24 \). The
stream function is dominated by a single, anti-clockwise cell. All
clockwise circulation is confined to two small regions close to the
upper and the lower boundary. Gradients in the salinity field
increase, and the wave pattern is skewed. On the other hand, because
the cell's turnover time is of the same order as the thermal
diffusivity time, the temperature field is only slightly perturbed.

At a later time, (\( t=48 \), fig.~\ref{fig:cycle}, third panel from
top), the large scale flow is close to its maximum intensity. All
clockwise circulation has disappeared from the stream function. A
sequence of wave roll-ups has made the salinity field homogeneous over
some parts of the domain, and it has greatly increased the gradients
elsewhere.  Notice the similarity between the stream function at this
time and the stream function of the steady solution. In both cases, the
circulation has the same sign in the whole domain, and the convective
cells are separated by saddle points. Here, however, the strength of
convection is weaker than in the steady solution. The flow is too
slow to advect the temperature field appreciably, and the salinity
plumes are too heavy to be sustained by the flow in a steady
equilibrium.

The sequence of wave roll-ups continues until the shear becomes too
intense to allow for any further roll-up ( \( t=63 \),
fig.~\ref{fig:cycle}, fourth panel from top). At that point, the high
shear disrupts the convective cells and destroys the inhomogeneities
in the temperature and salinity fields, bringing them close to the
conductive solution.

For the remaining fraction of the whole cycle, the flow is dominated
by an intense, \( x \)-independent shear which flows rightward in the
upper half of the domain, and leftward in the lower
half ($t=163$, fig.~\ref{fig:cycle}, fifth panel from top). As for the steady
 case, for each solution there is a
symmetric one with the shear flowing in the opposite direction; the
particular choice of initial conditions selects between the two
possibilities. The temperature and the salinity fields closely
resemble the conductive solution \( T=S=-z \).  The amplitudes \(
W^{+} \), \( W^{-} \), \( \Theta ^{+} \), \( \Theta ^{-} \), \( \Phi
^{+} \) and \( \Phi ^{-} \) are very small (less than \( 5\cdot
10^{-4} \) in our non-dimensional units) at any height \( z \). The
evolution of \( U \) is essentially decoupled from the dynamics of the
other variables, and it is reduced to a slow viscous decay. Finally,
when the intensity of the shear has been sufficiently damped, the
overstable oscillations restart their growth, and the cycle begins
again.

\begin{figure}
{\centering \resizebox*{0.7\textwidth}{!}
{\rotatebox{0}{\includegraphics{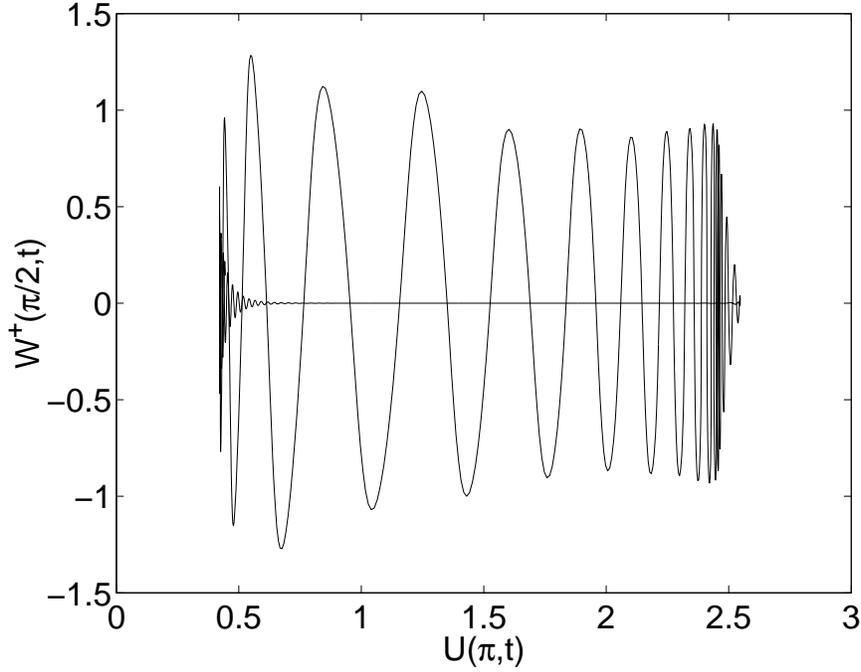}}} \par}

\caption{Shearing velocity $U$ at top boundary vs vertical
velocity amplitude $W^+$ at the middle of the computational domain
during the cycle of Fig. \ref{fig:Uz0}.\label{fig:UvsW}}
\end{figure}

The dynamics that we just described is summarized in
figure~\ref{fig:UvsW}, where we plot the value of the shearing
velocity $U$ at the top boundary versus the value of $W^+$ in the
middle of the slab.

With the parameters discussed here, the cycles are not perfectly
periodic. In fact, longer simulations reveal that the cyclic behavior
is only a long transient. After approximately 4500 thermal times the
flow falls back to the steady solution. On the other hand, at lower
Rayleigh numbers, such as $R_T=-100$ and $R_S=-200$, the wave-shear
mixing cycle appears to be a perfectly periodic solution. Further
exploration of the parameter space has shown that, either as a
transient or as the asymptotic regime, the roll-up of overstable
oscillations by a self-excited shear is a behavior characteristic of
the low Prandtl number regime.

\begin{figure}
{\centering \resizebox*{1\textwidth}{!}
{\rotatebox{-0}{\includegraphics{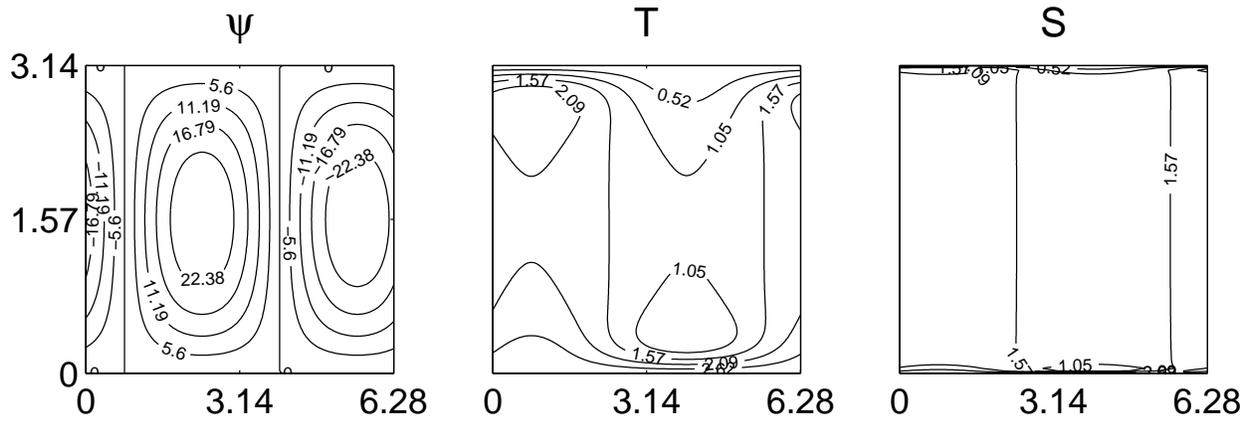}}} \par}

\caption{Stream function, temperature and salinity fields in the
simulation with \protect\( U\equiv 0\protect \).  This is a
steady solution. The fields are periodic in the horizontal. One period
is shown.\label{fig:noshear}}
\end{figure}

In order to bring out the key role of the large-scale flow, we have
carried out a simulation where the variable \( U \) was kept equal to
zero at all times, which is equivalent to removing the first of the
equations (\ref{eq:1mode_model}).  Starting from initial conditions of
small amplitude, the numerical solution undergoes an oscillatory
instability in agreement with the predictions of the linear
analysis. However, after about 100 thermal times the solution has
converged to a steady state, shown in figure~\ref{fig:noshear}. In
this solution, all the salinity gradients are concentrated at the
boundaries. In the interior of the domain the salinity is almost
constant. This brings the system to a state analogous to purely
thermal convection without shear, with two symmetric cells in the
stream function field, and two steady thermal plumes (one ascending,
the other descending) in steady equilibrium between advection and
diffusion.

\subsection{A Lagrangian view of the wave-shear mixing.\label{sec:lagrangian}}

The numerical simulations described above show that the
fluid can alternate long periods of laminar motion with relatively shorter
times of tumultuous overturning. To better investigate the details
of these \emph{mixing events} we have computed the trajectories of a
sample of fluid particles. For this purpose,
the variables \( U \),
\( W^{+} \) and \( W^{-} \) are interpolated using cubic spines
(taking care to use the correct boundary conditions in their spline
representation), to recover the fluid velocity \( \mathbf{u}_{i}(x,z) \)
at the position of the \( i \)-th particle. The advection equation
\( \dot{\mathbf{x}}_{i}=\mathbf{u}_{i} \) is then integrated at each
Eulerian time step with a second order Adams-Bashfort scheme to obtain
the particles' trajectories.

\begin{figure}
{\centering \resizebox*{0.8\textwidth}{!}
{\rotatebox{0}{\includegraphics{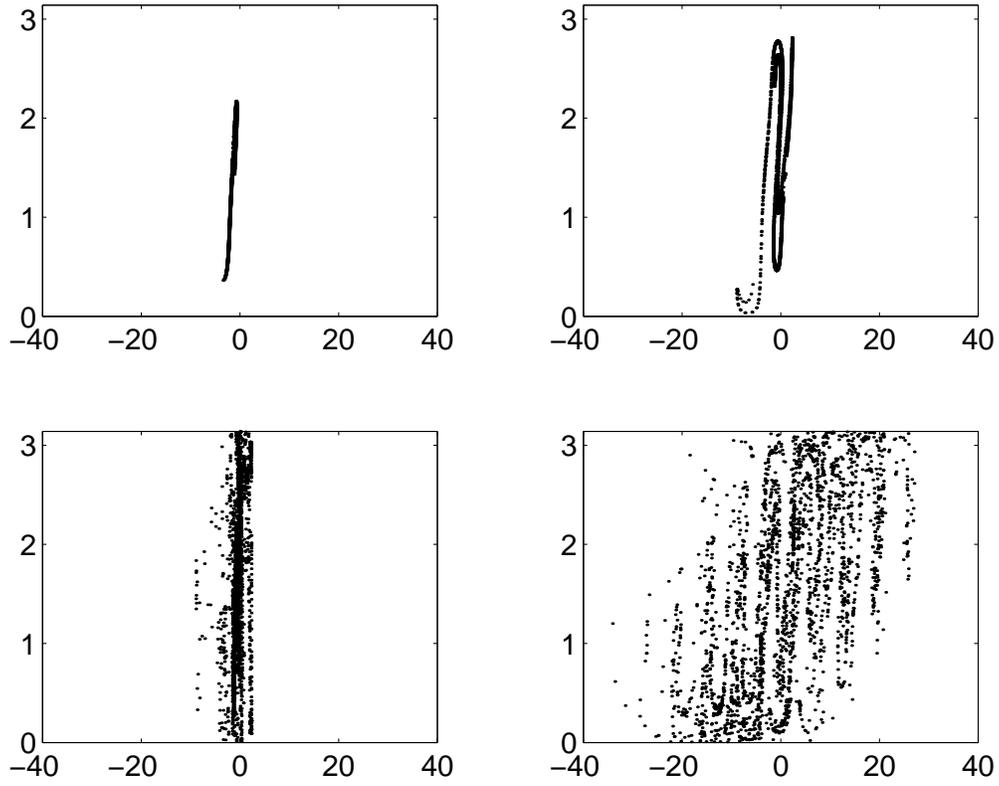}}} \par}

\caption{Positions of 2500 Lagrangian tracers at different times. From
left to right, top to bottom: \protect\( t=15\protect \); \protect\(
t=20\protect \); \protect\( t=25\protect \); \protect\( t=30\protect
\).  The tracers were released at time \protect\( t=0 \protect \)
({\em cf.} Fig. \ref{fig:Uz0}).
\label{fig:particles_snaps}}
\end{figure}

We release 2500 particles at time $t=0$ on a square array of 50 by 50
points, with side length of 0.5 non-dimensional units, and with the
lower left corner at (0, \( \pi /3 \)).  As time goes on, the first
part of each thermohaline oscillation stretches the
ensemble of particles vertically, the large-scale flow slants it, and the second
part of the oscillation folds it. The interplay of waves and shear
produces a dynamics very similar to that of a Baker's map, and it is
pictured in the time sequence of figure~\ref{fig:particles_snaps} (for
an introduction to the physics of mixing see Ottino (1989); a simple
example of such mixing flow is given by Aref's blinking vortex,
Aref, 1984).  The stretching and folding continues until the
thermohaline oscillations are disrupted by the shearing velocity. By
the time $U$ reaches its maximum amplitude, ($t \simeq 65$), the
distribution of tracers has become vertically homogeneous.

\begin{figure}
{\centering \resizebox*{0.7\textwidth}{!}
{\rotatebox{0}{\includegraphics{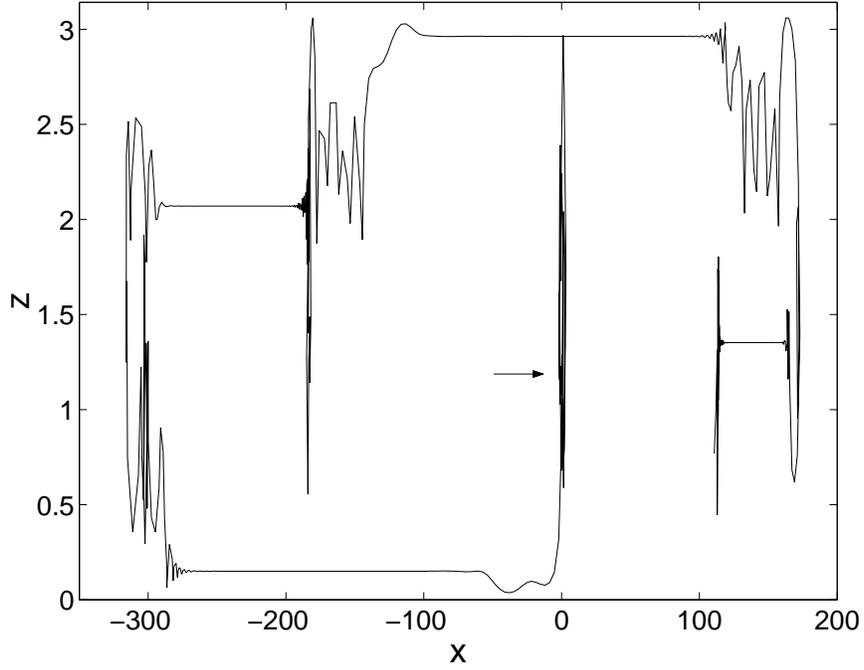}}} \par}

\caption{Trajectory of a single Lagrangian tracer. The path begins at
the position marked by the arrow. The integration lasts for 1000
thermal times. Four mixing events, and the beginning of a fifth, are
clearly visible, alternated with periods where the advection is only
along the \protect\( x\protect \)-axis. \label{fig:tracer1075}}
\end{figure}

The trajectory of an individual Lagrangian tracer ({\em cf.}
fig.~\ref{fig:tracer1075} ) illustrates what happens on longer time
scales.  The mixing events are clearly singled out by the periods of
wiggly motion, alternating with periods where the advection is due to 
only the large scale velocity \( U \). Since the sign of the shear
remains constant, when a Lagrangian tracer happens to be in the upper
half of the computational domain, it travels rightward, while it
travels leftward when in the lower half.  Dispersion
in the horizontal direction occurs because the \( z \) coordinate of a
particle after a mixing event has a sensitive dependence upon the
particle's position before the mixing event. Each particle switches in
an unpredictable way between rightward and 
leftward motion, performing a random walk along the \( x \)-axis.
Thus we expect the horizontal transport of particles to be described
by a diffusion process. On a dimensional basis, we can express its
diffusion coefficient as $K=U_c^2 t_c$. We estimate $U_c$ as the time
and space average of $|U|$ and $t_c$ as the duration of the viscous
decay of the shearing velocity that separates a mixing event from the
next. In our non-dimensional units we have $U_c \approx 0.85$, $t_c
\approx 180$, which yields $K \approx 130$.

This qualitative description is confirmed by evaluating the single-particle
dispersion $\xi^2$
\begin{equation}
\label{eq:single-part-dispersion}
\xi^2(t;\mathbf{x}_i(t_{0}))=\frac{1}{N}\sum _{i=1}^{N}\left|
\mathbf{x}_{i}(t)-\mathbf{x}_{i}(t_{0})\right| ^{2}
\end{equation}
where \( N \) is the number of Lagrangian tracers and \(
\mathbf{x}_{i}(t_{o}) \) is the initial position of the \( i \)-th
one. In the case of Brownian motion, \( \xi^2 \) does not depend on
the initial positions and it is related to the diffusion coefficient
\( K \) of the Brownian stochastic process by the formula \(
\xi^2(t)=2Kt \) (Gardiner, 1996). In our case, since the Eulerian flow has
very different characteristics along the \( x \) and the \( z \)
direction, we apply the definition of single-particle dispersion
separately to each direction.

\begin{figure}
{\centering
\resizebox*{0.45\textwidth}{!}
{\rotatebox{0}{\includegraphics{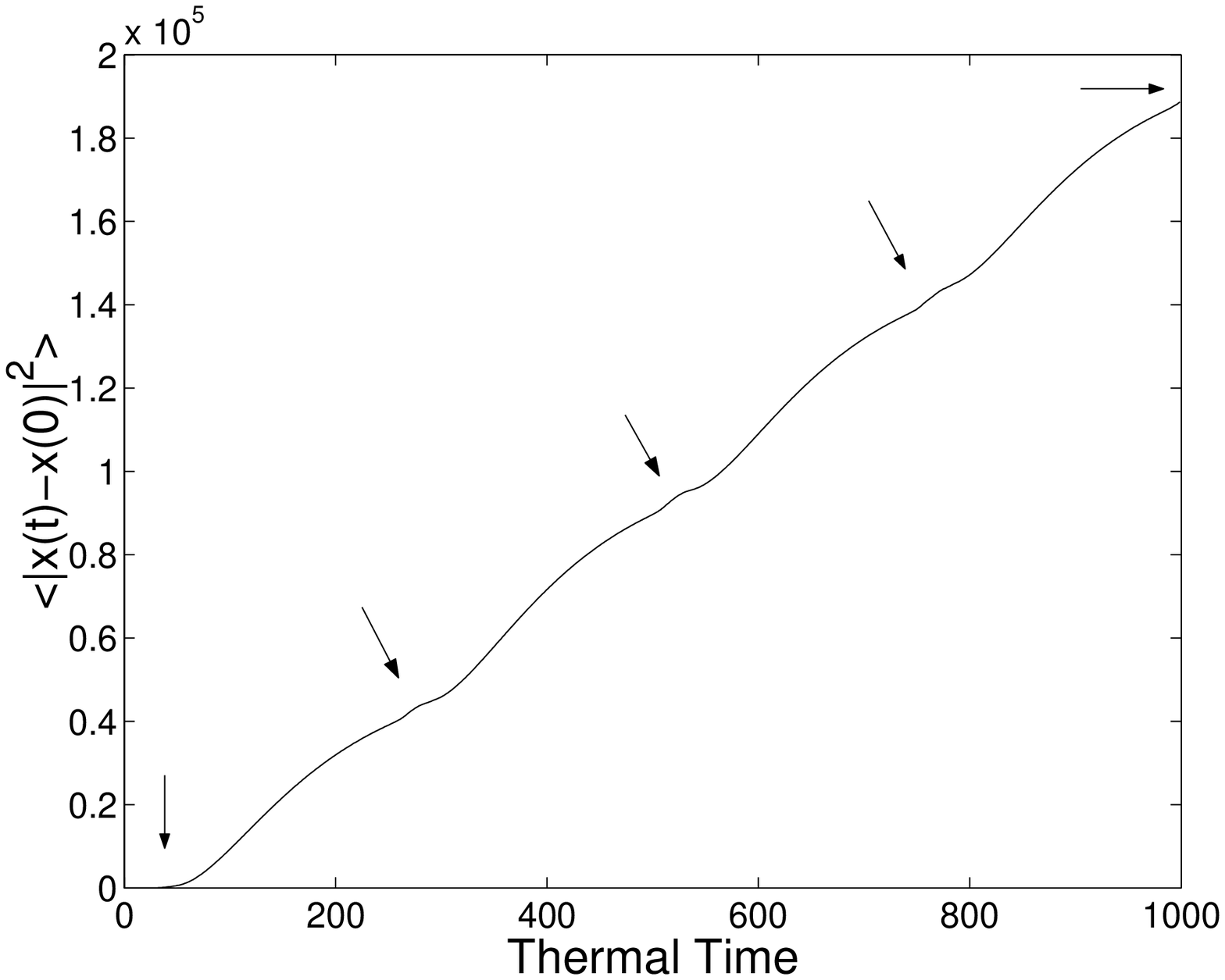}}}
\resizebox*{0.45\textwidth}{!}
{\rotatebox{0}{\includegraphics{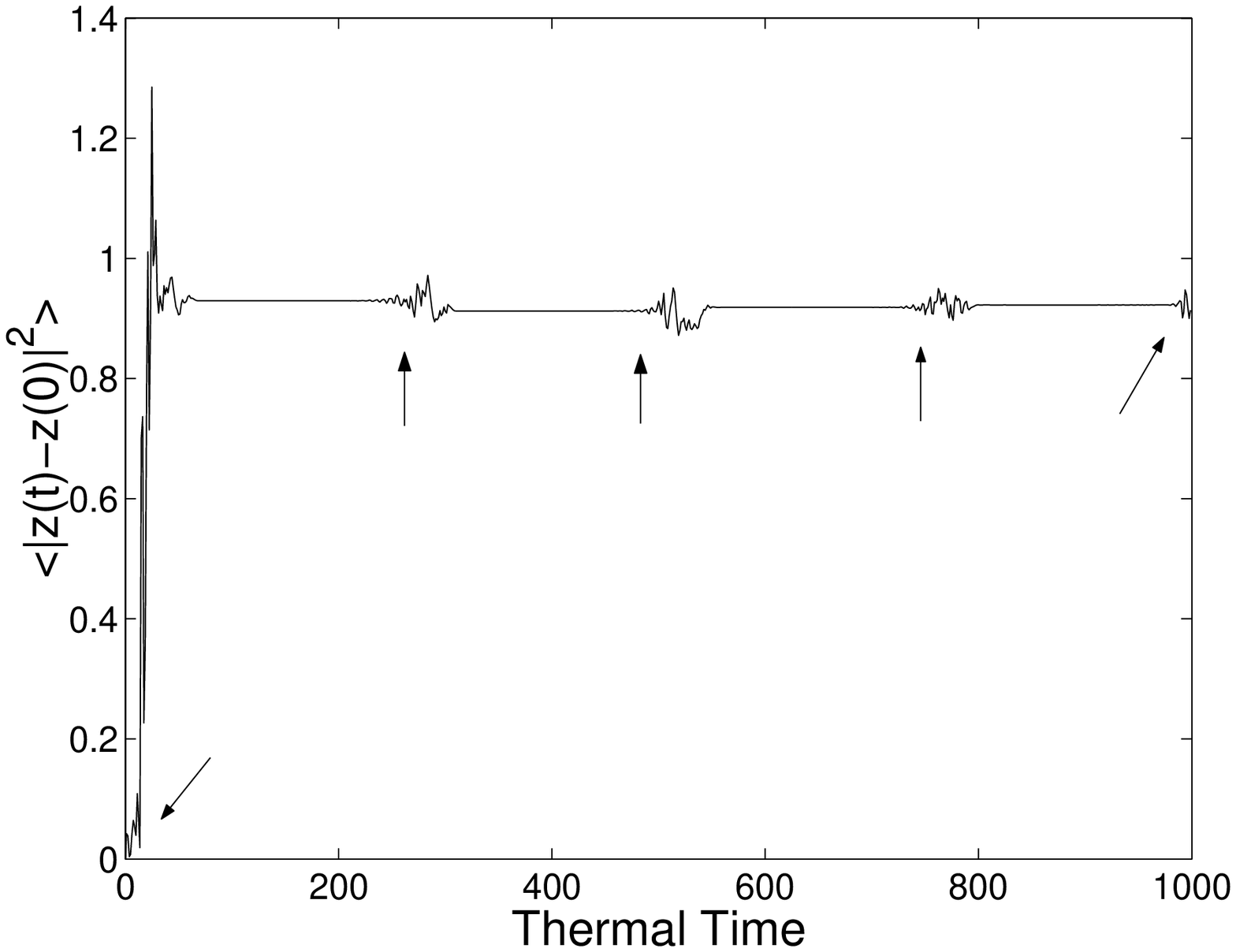}}}
}

\caption{Single-particle dispersion along the horizontal (left panel) and
the vertical (right panel). The times when sequences of wave roll-ups
(mixing events) occur are marked by the arrows.\label{fig:single-part-disp}}
\end{figure}

The results are shown in figure~\ref{fig:single-part-disp}. The overall
trend of \( \xi^2(t) \) for horizontal displacements is well
approximated by a linear fit, which gives an effective diffusion
coefficient \( K=98.6\pm 1.0 \) in non-dimensional units, in good
agreement with the above rough estimate. However, on short time scales,
the behavior of \( \xi^{2}(t) \) is not perfectly linear.  Initially,
in the \( x \)-direction, dispersion grows slowly, because the tracers
are all clustered together into a single, coherent convection
cell. After the first vertical mixing event (marked by the arrow in
figure~\ref{fig:single-part-disp}), the horizontal dispersion grows
faster with time. At this stage, the dispersion is approximately
ballistic (that is \( \xi^2 \propto t^{2} \)) until \( U \) can be
approximated by a constant. However, as the large scale flow \( U \)
decays, the dispersion slows down as well, until the next mixing event
occurs.

Along the vertical direction, the single-particle dispersion grows
from zero to its maximum value during the first occurrence of wave-rollups,
which makes the tracers' distribution homogeneous in the vertical.
The vertical dispersion then remains approximately constant, with
statistical fluctuations during the vertical mixing events.

To confirm that the flow of figure~\ref{fig:cycle} produces authentic
mixing, we look for evidence that particle paths are independent of
each other by evaluating the pair dispersion $\Xi^2$
\begin{equation}
\label{eq:pair-dispersion}
\Xi^2(t)=\frac{2}{N(N-1)}\sum _{i=1}^{N} \sum_{j>i}
\left| \mathbf{x}_{i}(t)-\mathbf{x}_{j}(t)\right| ^{2}.
\end{equation}
If each particle follows an independent Brownian path, then
$\Xi(t)=4Kt$. On evaluating the diffusion coefficient along the
$x$-direction by using pair dispersion, we find $K=102.0 \pm
1.0$, in close agreement with the value obtained in the
single particle analysis. This is strong evidence that the long-time transport
properties of the flow in the horizontal direction are those of 
Brownian motion, and can thus be modeled by a suitable eddy diffusion
coefficient.

We carried out a similar simulation for the case where the large-scale
flow is not allowed. As expected when the Eulerian flow is steady,
the tracers moved along periodic orbits. Even after 1000 thermal times
the 2500 tracers, seeded at the same initial position as the tracers
in the simulation described above, remain confined in a circular belt
and do not disperse.

\begin{figure}
{\centering \resizebox*{1\textwidth}{!}{\rotatebox{-0}
{\includegraphics{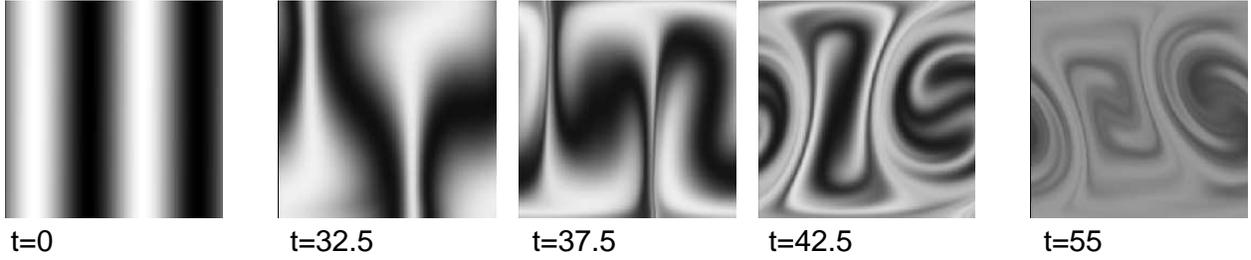}}} \par}

\caption{Evolution of the concentration of a passive scalar during a
complete cycle of the flow in figure~\ref{fig:cycle}.  The grey scale
is between 0 and 1.  The transition from oscillation to direct
convection is visible in the 3 central panels whereas the last one
shows how much homogenization occurred during one cycle. The fields are
periodic in the horizontal. One period is shown.
\label{fig:pas_scalar}}
\end{figure}

To complement the information given by the advection of individual
Lagrangian tracers, and to give a further visual indication that no
barriers to mixing exist in the flow of figure~\ref{fig:cycle}, we
carried out some simulations in which we use the streamfunction
generated by our model to advect a passive scalar field $c$, according
to the equation
\begin{equation}
\partial _{t} c - J(\psi, c)= \frac{\kappa_c}{\kappa_T}\nabla^2 c.
\label{eq:advection_diffusion}
\end{equation}
The small amount of diffusion $\frac{\kappa_c}{\kappa_T}=0.001$ is
necessary in order to maintain the stability of the finite volume code
used to integrate equation
(\ref{eq:advection_diffusion}). Figure~\ref{fig:pas_scalar} shows a
time sequence of the numerical solution, and illustrates the high
degree of homogeneity reached by the passive scalar during a single
wave-shear mixing cycle.

\section{ODD convection with heat and salt.}

Because conditions favoring the onset of ODD convection occur at several
places in the Earth's oceans
we investigate the behavior of the model in the parameter range
appropriate for sea water.   For this, we adopt
\( \sigma=10 \), and \( \tau=10^{2} \), which are approximately the
values of the Prandtl and the Lewis numbers of sea water. With these
parameters, the ODD instability occurs when the density ratio
\( R_{\rho }={R_{T}}/{R_{S}}\) approaches unity (Baines \& Gill, 1969).

\begin{figure}
{\centering \resizebox*{1\textwidth}{!}{\rotatebox{-0}
{\includegraphics{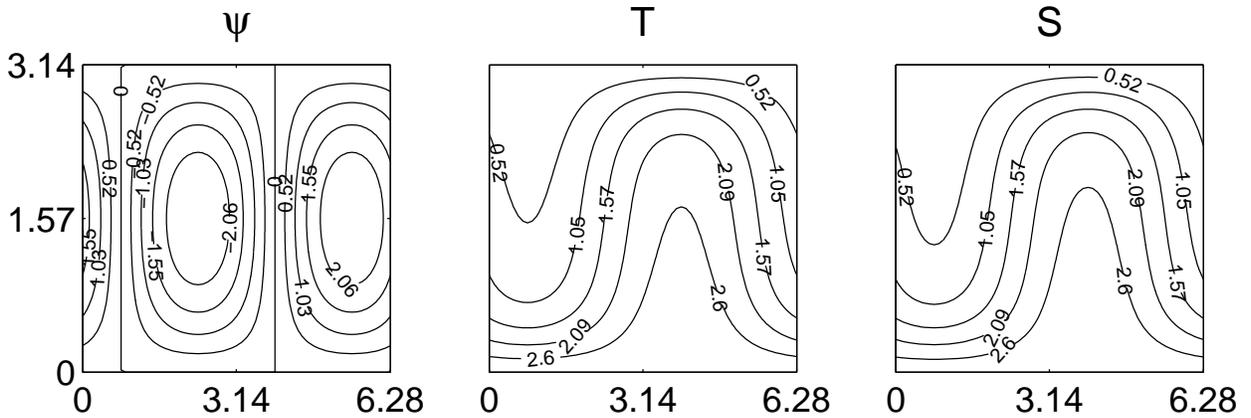}}} \par}

\caption{Stream function, temperature and salinity fields for the heat
and salt simulation with \protect\( R_T=-200 \protect \) and
\protect\(R_{\rho} = 0.99\protect \).  The fields are periodic in the
horizontal.  One period is shown.\label{fig:ra200}}
\end{figure}

When \( R_{T}=-200 \), \( R_{\rho }=0.99 \) the system converges to a
limit cycle characterized by a pair of plumes which oscillate up and
down as standing waves, with periods of $\simeq 0.61$ thermal times.
Figure~\ref{fig:ra200} shows a snapshot of the stream function,
temperature and salinity fields close to the extremum of one
oscillation: the warm, salty plume on the right has just reached the
maximum elevation, the cold, fresh plume on the left has reached
its maximum depth, and they are both moving back toward the middle of
the domain. The large scale flow is not excited, and initial
perturbations on \( U \) decay as \( e^{-\sigma t} \). No Lagrangian
mixing occurs in this simple flow.

Searching for a self-excited large-scale flow, we increased the
Rayleigh number up to \( R_{T}=-1000 \), and we have set
the density ratio to \( R_{\rho }=0.95 \). To accurately resolve the
high salinity gradients that develop at the top and bottom boundaries,
we increased the vertical resolution to 256 modes. The initial conditions are
the same as in section \ref{sec:wave-shear}. We have tried several
values for the initial amplitude \( A \) in the range
\( [10^{-3},10] \). The solution that we find exhibits extremely long
cycles of approximately \( 100 \) thermal times, independently of the
value of \( A \).

\begin{figure}
{\centering \resizebox*{1\textwidth}{!}{\rotatebox{-0}
{\includegraphics{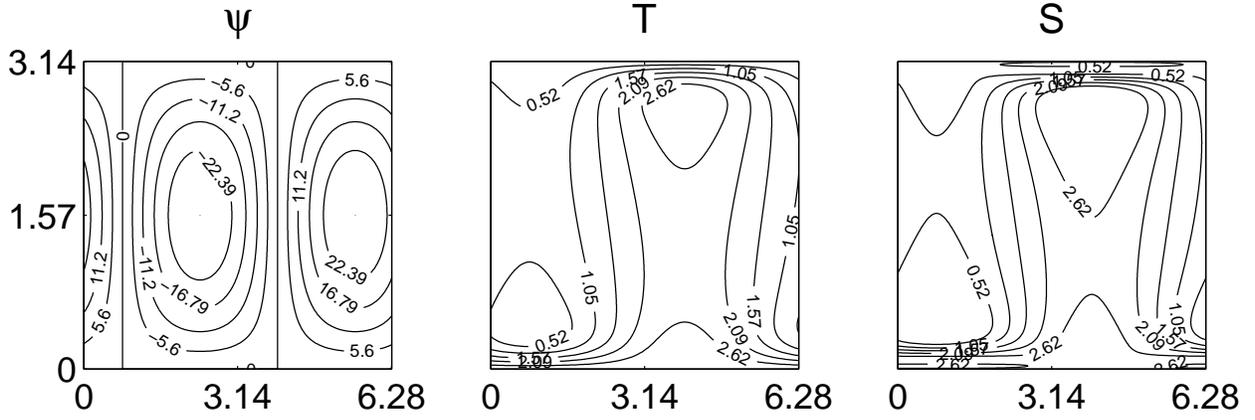}}} \par}

\caption{Streamfunction, temperature and salinity fields for the heat
and salt simulation with \protect\( R_T=-1000\protect \) at time
\protect\( t=4.3\protect \). The phase of the oscillation is the same as in figure~\ref{fig:ra200} The fields are periodic in the
horizontal. One period is shown.\label{fig:ra1000} }
\end{figure}

Initially this solution shows a pair of oscillating plumes analogous
to those found at \( R_{T}=-200 \). They move faster (the period of
oscillation is 0.3 thermal times, but it shrinks as the amplitude of
the oscillations grows), and lead to sharper gradients in the scalar
fields. The large-scale flow is excited, but it is so weak that the
associate skewness of the convective cells is imperceptible in the
figures. Figure~\ref{fig:ra1000} shows the stream function, the temperature
and the salinity fields for the simulation starting with \( A=0.01 \)
at time \( t=4.3 \), when the salinity fluctuations reach their
maximum amplitude. The oscillations, however, do not last indefinitely
as in the previous case. They damp out in a few thermal times, and the
motion in the fluid ceases almost completely.

\begin{figure}
{\centering \resizebox*{0.7\textwidth}{!}
{\rotatebox{0}{\includegraphics{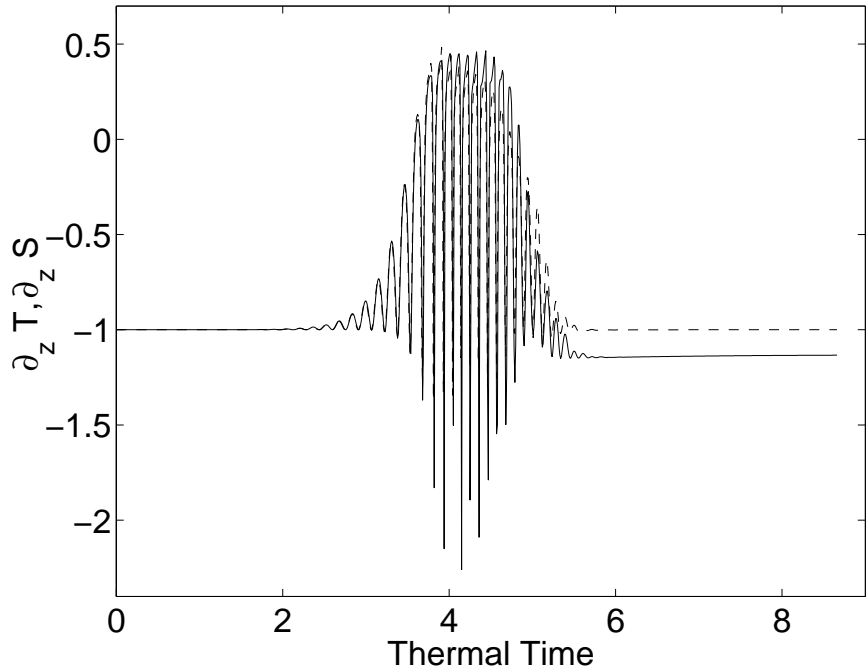}}} \par}

\caption{Time evolution of \protect\( \partial _{z}\overline{T}\protect \)
(dashed line) and \protect\( \partial _{z}\overline{S}\protect \)
(solid line) computed at \protect\( z=\pi /2\protect \) in the simulation
with \protect\( R_T=-1000\protect \)
.\label{fig:dz}}
\end{figure}
\begin{figure}
{\centering \resizebox*{0.7\textwidth}{!}
{\rotatebox{0}{\includegraphics{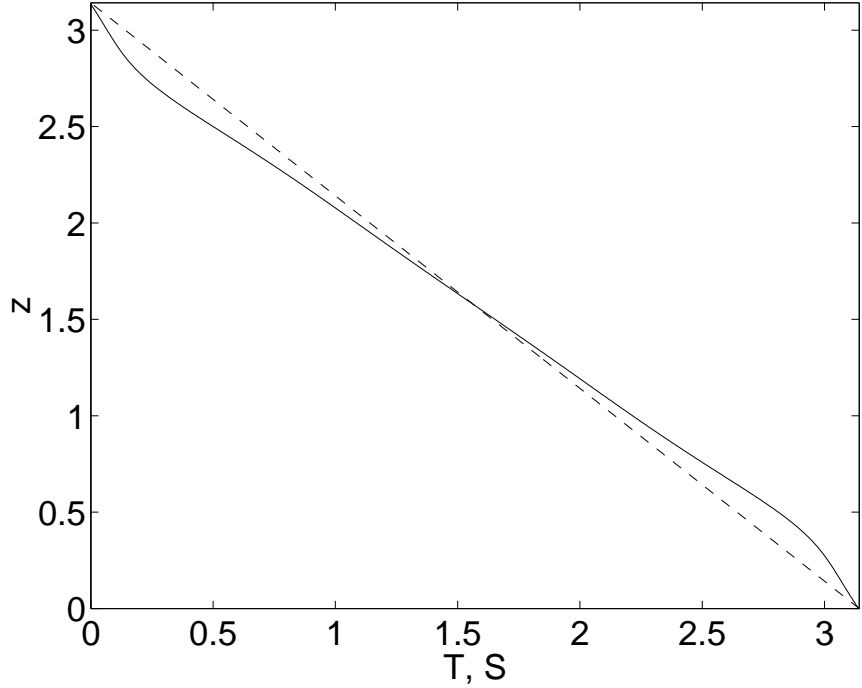}}} \par}

\caption{Horizontally averaged temperature (dashed line) and salinity (solid
line) in the simulation with
\protect\( R_T=-1000\protect \)
at the time \protect\( t=4\protect \).
\label{fig:ra1000TS}}
\end{figure}

By looking at the vertical profiles of temperature and salinity it is
possible to find a hint of the mechanism that switches off the
oscillations.  Figure~\ref{fig:dz} shows the time evolution of
\( \partial_{z} \overline{T} \) and \( \partial_{z}\overline{S} \)
computed at \( z=\pi /2 \) (that is, in the middle of the slab).
Initially \( \partial_{z}\overline{T}=\partial_{z}\overline{S}=-1 \).
As the doubly-diffusive oscillations grow and then damp out, the slope
of the salinity profile becomes steeper than the slope of temperature, and
one is left with \( \partial_{z}\overline{S}<-1 \), while the
temperature slope promptly returns to the diffusive solution \(
\partial_{z}\overline{T}=-1 \).  This results in a decreased density
ratio in the middle of the slab, which locally makes the fluid
linearly stable. Figure~\ref{fig:ra1000TS} shows the horizontally
averaged profiles of temperature and salinity at the time
\( t=6 \). The density ratio at \( z=\pi /2 \) is now
\( R_{\rho }=\frac{\partial_{z}\overline{T}}
{\partial_{z} \overline{S}}R_{\rho _{\rm initial}}=0.831 \), which is less
than the critical value \( R_{\rho _{\rm critical}}=0.9163\ldots \) required
to have a linear instability for the chosen values of \( R_{T} \), \(
\sigma \), and \( \tau \). At this time all motion has become negligibly small.
The only dynamic in the system is the slow diffusion of the salinity
field, which relaxes to the conductive solution with a time scale which,
in non dimensional units is \( O(\tau) \). Finally, when \( t\approx 100 \),
the growth of ODD instability resumes and the whole cycle starts over
with an alternation of short-lived oscillatory convection and slow
diffusive relaxation. This dynamics does not lead to mixing. A set of
Lagrangian tracers seeded in the way described in section
\ref{sec:lagrangian} remains closely packed during the whole
duration of the oscillations.

While the salinity build-up stabilizes the bulk of the fluid, we
observe in figure~\ref{fig:ra1000TS} that, near the boundaries,
\( \partial_{z}\overline{T}<\partial_{z}\overline{S}\),
implying that the fluid there is unstably stratified. However, we
do not observe direct convection being triggered at the boundaries.
We assume that this is a consequence of the truncation of the Boussinesq
equations (\ref{2Dboussinesq}): the range of horizontal wavenumbers
which would be unstable is not represented in the truncated model
(\ref{eq:1mode_model}). Numerical solutions of the untruncated equations
are needed in order to show whether convection actually arises at the
boundaries and in what ways it contributes to the establishment of the
step-like profiles seen in laboratory experiments (Huppert \& Linden, 1979).

In this simulation the large-scale flow \( U \) does not seem to play
a dominant role. The doubly-diffusive oscillations are suppressed
by the salinity build-up in the center of the domain, and not by the
shear induced by \( U \), as in the low Prandtl number case. Indeed, a
simulation with \( R_{T}=-500 \) and \( R_{\rho }=0.95 \) shows the
same qualitative dynamics as the simulations with \( R_{T}=-1000 \),
but there is no self-excitation of a large-scale flow. 

\begin{figure}
{\centering \resizebox*{1\textwidth}{!}{\rotatebox{-0}
{\includegraphics{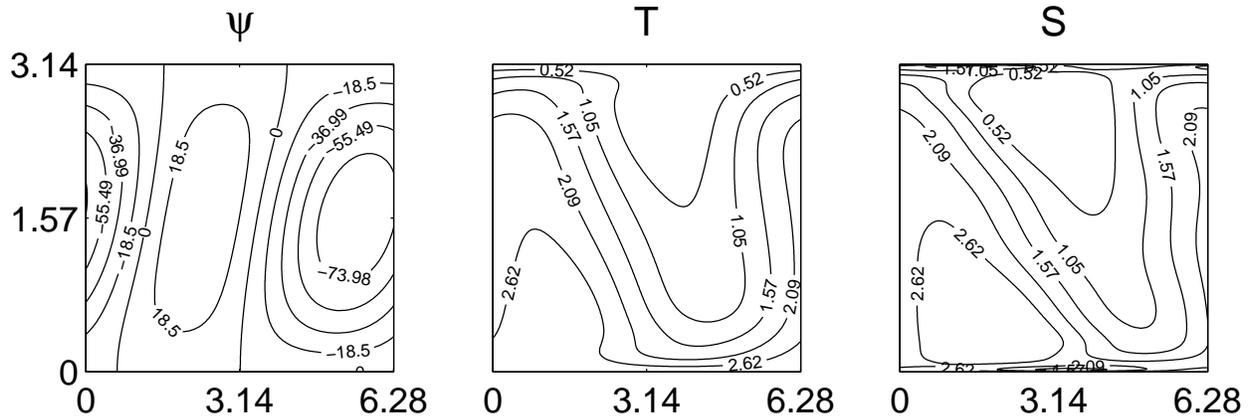}}} \par}

\caption{Streamfunction, temperature and salinity fields at time
\protect\( t=4.3\protect \) for the heat
and salt simulation with \protect\( R_T=-1000\protect \) and an
externally imposed shear $\omega _{0}=0.2$. The fields are periodic in the
horizontal. One period is shown.\label{fig:sheared_sea} }
\end{figure}

\begin{figure}
{\centering \resizebox*{0.8\textwidth}{!}
{\rotatebox{0}{\includegraphics{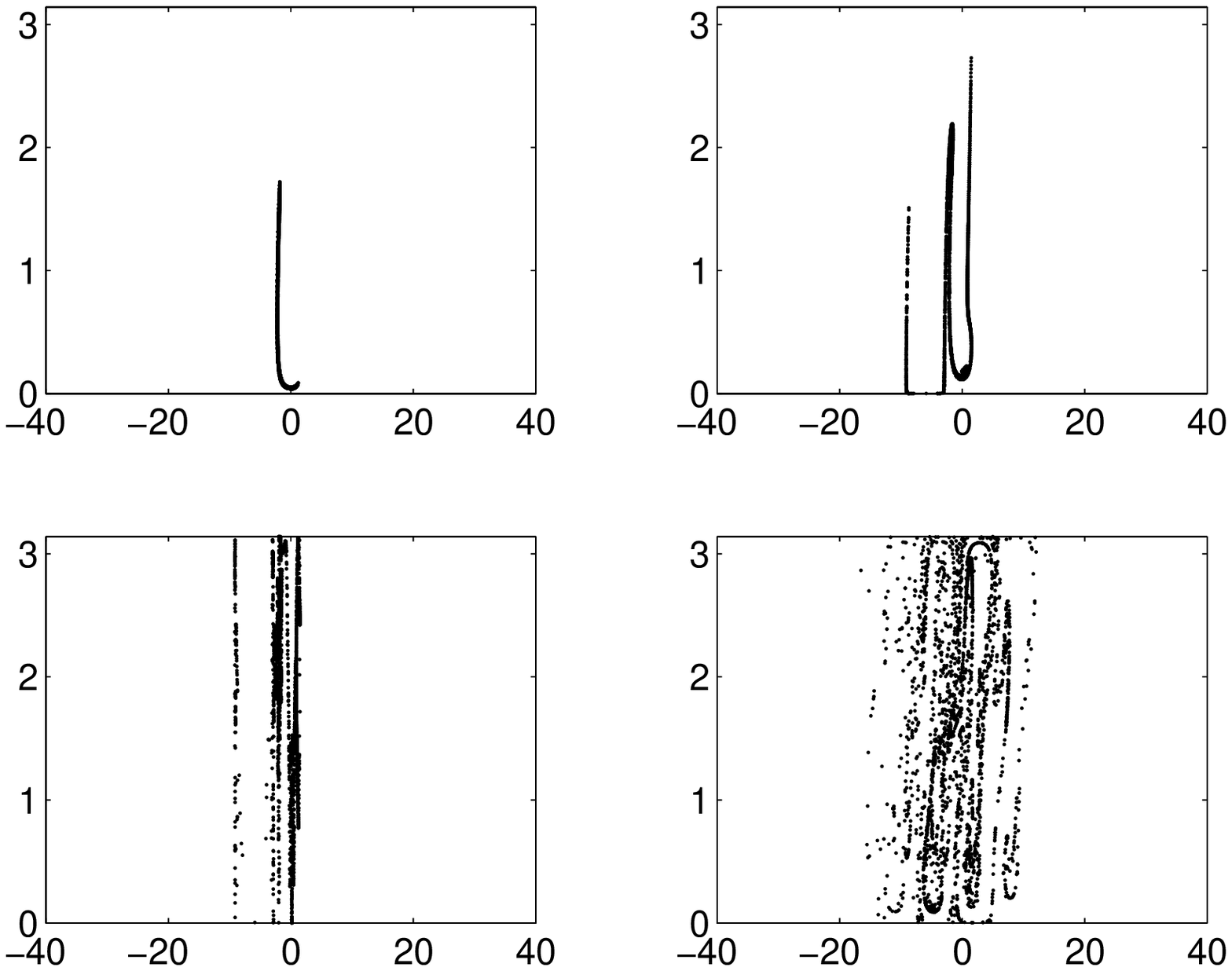}}} \par}

\caption{Positions of 2500 Lagrangian tracers at different times in a
simulation with sea water parameters and an externally imposed
shear. From left to right, top to bottom: \protect\( t=4\protect \);
\protect\( t=4.15\protect \); \protect\( t=4.3\protect \); \protect\(
t=4.45\protect \).
\label{fig:seaw_streaks}}
\end{figure}

However, large-scale flows are prevalent in geophysics where they need
not be convectively driven.  Such flows may play the kind of role in
the convective process that the shears we have been discussing do.
For this reason, we have investigated the interaction between ODD
oscillations and an imposed shearing velocity,
\( U=\widetilde{U}+\omega _{0}z \) in the equations
(\ref{eq:1mode_model}), and solving for \( \widetilde{U} \).  Here \(
\omega _{0} \) is a constant shear. In our spectral code we represent
\( z \) as a trigonometric sum in the interval \( [0,\pi ] \),
truncated at the Nyquist frequency.  We keep \( R_{T}=-1000 \), \(
R_{\rho }=0.95 \), and we impose an external shear
\( \omega _{0}=0.2 \).
Once again, oscillating plumes develop as a result of the ODD
instability. The presence of the external shear makes the convective
cells sufficiently asymmetric that they now can contribute to the
build-up of a large-scale flow, which can reach a peak intensity of
$30$ non dimensional units close to the boundaries. A snapshot of the
fields at time $t=4.3$ is shown in figure~\ref{fig:sheared_sea}.
As in the case with low Prandtl number, the large-scale flow induces
wave roll-ups, which mix the fluid until the salinity build-up chokes
the motion. The resulting Lagrangian chaos is illustrated in
figure~\ref{fig:seaw_streaks}.

\section{Discussion and conclusions}

This exploratory study of sheared ODD convection has revealed some
previously unknown phenomenology. The main finding is that the
presence of a large-scale shearing velocity (be it self-excited or
externally imposed) induces vigorous mixing. The mechanism is
reminiscent of the mixing induced by the Baker's map: ODD waves
vertically stretch material lines, the shear slants them, so that they
can subsequently be folded up. This engenders a rapid, effective
homogenization of Lagrangian tracers along the vertical.  Horizontal
mixing is also enhanced, in a way which recalls the phenomenon of
shear dispersion (see, e.g. Jones \& Young, 1994). As first
noticed by Taylor, in the presence of a shear, molecular
diffusion is enhanced along the direction of the flow. In our model
the wave-shear mixing plays the role of molecular diffusion.

Further work is needed to asses at which Rayleigh numbers the large-scale flow
appears.  Spontaneous generation of shear seems very elusive at high
Prandtl number, as in the case of heat and salt in water.  This is
supported by the laboratory experiments of Stamp \& Griffiths (1997)
and Stamp et al. (1998). They observe the generation of
large-scale flows in convecting regions above and below an interface
having ODD-favorable stratification (therefore giving further
indication that this is a common feature of convecting fluids). The
flows have the same intensity and direction above and below the
interface, so they do not cause any shear on the interface, but merely
translate it, either leftward or rightward. In an annular geometry, in
the reference frame comoving with the mean flow, they observe gross
features in agreement with the results on step-like interfaces of
Gough \& Toomre (1982), who employ a truncated model similar to ours.

On the other hand, we find that large-scale shearing flow is easily
generated at very low Prandtl number, which is of interest in
astrophysical applications. Unfortunately, carrying out either
laboratory experiments or reliable direct numerical simulations in
this regime is challenging.
Other mechanisms can lead to mixing, distinct from that illustrated in
this work. In particular, the simulations of Merryfield (1995) and of
Biello (2001) both show ODD oscillations breaking into small-scale
turbulence, as in the earlier discussion of Stevenson (1979).  A
larger investigation is needed to decide which of these mechanisms may
be the more significant in different contexts. 

In all cases where we have found mixing-inducing motion, it lasts for a
relatively short interval, and it is then followed by a longer interval of
non-mixing dynamics. The physical cause of this intermittent
behavior is different in the low Prandtl number regime and in the
sea water regime. In the first case the stabilizing factor is the shearing
velocity \( U \); in the second case the fluid is made linearly stable by
a salinity build-up away from the boundaries.
The present study has suggested that the astrophysical situation is the
one in which the convectively induced shears come into play.  This points
to the value of a study directed at low Prandtl number ODD convection and
suggests that we next seek simplifications that may help us in that limit.

\acknowledgments
{\noindent} {\bf Acknowledgments}  ~
We thank the GFD program for bringing us all together. E. A. Spiegel
was supported by NSF Applied Mathematics under DMS-99-72022. S. Talon 
was supported by  NSERC of Canada and by the Canada Research Chair
in Stellar Astrophysics awarded to Gilles Fontaine.
We are grateful to Barry Ruddick for his comments which contributed to
improve this
manuscript and to Bill Young who helped us clarifying the issue of
what defines mixing.

\newpage

\thispagestyle{empty}

\begin{thebibliography}{26}

\bibitem[]{Aref84}
Aref, H.
``Stirring by chaotic advection'', \emph{J. Fluid
Mech.}, \textbf{143}, 1, (1984).

\bibitem[]{Baines69}
Baines, P. G. and Gill, A. E.,
``On thermohaline convection with linear gradients'', \emph{J. Fluid
Mech.}, \textbf{37}, 289, (1969).

\bibitem[]{Biello01}
Biello, J. A.,
``Layer formation in semiconvection'', preprint: astro-ph/0102338,
http://arXiv.org/abs/astro-ph/0102338.

\bibitem[]{DaCosta81}
Da Costa, L.N., Knobloch, E. and Weiss, N.O., ``Oscillations in
double-diffusive convection'', \emph{J. Fluid Mech.},  \textbf{109},
 25-43 (1981).

\bibitem[]{Franceschini85}Franceschini, V. and Tebaldi, C.,
``Truncations to 12, 14 and 18 modes of the Navier-Stokes equations on
a two-dimensional torus'',
\emph{Meccanica}, \textbf{20}, 207, (1985).

\bibitem[]{Gardiner96}Gardiner, C. W.,
``Handbook of Stochastic Methods'', II edition, Springer-Verlag, Berlin
(1996).

\bibitem[]{Gough75}Gough, D. O., Spiegel, E. A. and Toomre, J.,
``Modal equations for cellular convection'',
\emph{J.Fluid Mech.}, \textbf{68}, 695, (1975).

\bibitem[]{Gough82}Gough, D. O. and Toomre, J.,
``Single-mode theory of diffusive layers in thermohaline convection'',
\emph{J.Fluid Mech.}, \textbf{125}, 75, (1982).

\bibitem[]{Howard86}Howard, L. N., and Krishnamurti, R.,
``Large scale flow in turbulent convection: a mathematical model'',
\emph{J. Fluid Mech.},
\textbf{170},385, (1986).

\bibitem[]{Huppert79}Huppert, H. E. and Linden, P. F.,
``On heating a stable salinity gradient from below'',
\emph{J. Fluid Mech.}, \textbf{95}, 431, (1979).

\bibitem[]{Jacobs81}Jacobs, C. A., Huppert, H. E., Holdsworth, G. and Drewy, D. J.,
``Thermohaline steps induced by melting at the Erebus Glacier tongue'', \emph{J. Geophys. Res.}, \textbf{86}, 6547, (1981).

\bibitem[]{Jones94}Jones, S. W. and Young, W.R.,
``Shear dispersion and anomalous diffusion by chaotic advection'',
\emph{J. Fluid Mech.},\textbf{280}, 149, (1994).

\bibitem[]{Kolodner90}Kolodner, P. R., Glazier, J. A. and Williams, H. L., ``Dispersive Chaos in One-Dimensional Traveling-Wave Convection'',
\emph{Phys. Rev. Lett.}, \textbf{65}, 1579, (1990).

\bibitem[]{Krishnamurti81}Krishnamurti, R. and Howard, L. N.,
``Large scale flow in turbulent convection'',
\emph{Proc. Natl. Acad. Sci. USA}, \textbf{78}, 1981, (1981).

\bibitem[]{Marmorino91}Marmorino, G. O., ``Intrusions and diffusive
interfaces in a salt fingering staircase'', \emph{Deep-Sea Research},
\textbf{38}, 1431, (1991).

\bibitem[]{Merryfield95}Merryfield, W. J.,
``Hydrodynamics of Semiconvection'', \emph{Astrophys. J.}, \textbf{444},
318, (1995).

\bibitem[]{Neal69}Neal, V. T., Neshiba, S. and Denner, W.,
``Thermal stratification in the Arctic Ocean'', \emph{Science}, \textbf{166}, 373, (1969).

\bibitem[]{Ottino}Ottino, J. M., 
``The Kinematics of Mixing: Stretching, chaos, and transport'', Cambridge University Press,
Cambridge (1989).

\bibitem[]{Pap99}Paparella, F. and Spiegel, E. A.,
``Sheared salt fingers: Instability in a truncated system'', \emph{Phys.
Fluids}, \textbf{11}, 1161, (1999). (PS)

\bibitem[]{Ruddick92}Ruddick, B., ``Intrusive mixing in a Mediterranean salt lens -- Intrusion slopes and dynamical mechanisms'',
\emph{J. Phys. Oceanogr.}, \textbf{22}, 1274, (1992).

\bibitem[]{Siggia94}
Siggia, E. D.,
``High Rayleigh Number Convection'', \emph{Annu. Rev. Fluid Mech},
\textbf{26}, 137, (1994).

\bibitem[]{Spiegel69}
Spiegel, E. A., ``Semiconvection'', \emph{Comments on Ap. and Space
Physics}, \textbf{1}, 57, (1969).

\bibitem[]{Stamp98}
Stamp, A. P., Hughes, G. O., Nokes, R. I. and Griffiths, R. W., ``The
coupling of waves and convection'', \emph{J. Fluid Mech.},
\textbf{372}, 231, (1998).

\bibitem[]{Stamp97}
Stamp, A. P. and Griffiths, R. W., ``Turbulent traveling-wave
convection in a two-layer system'', \emph{Phys. Fluids},
\textbf{9}, 963, (1997).

\bibitem[]{Stevenson79}
Stevenson, D. J., ``Semiconvection as the occasional breaking of weakly 
amplified internal waves'', \emph{Monthly Not. Royal Astr. Soc.},
\textbf{187}, 129, (1979).

\bibitem[]{Turner79}Turner, J. S.,
``Buoyancy effects in fluids'', Cambridge University Press, Cambridge
(1979).

\bibitem[]{Veronis65}Veronis, G.,
``On finite amplitude instability in thermohaline convection'', \emph{J.
Marine Res.}, \textbf{23}, 1, (1965).

\end{thebibliography}
\end{document}